\documentclass[reprint,aps,pre,twocolumn,notitlepage,showpacs,nofootinbib,preprintnumbers,superscriptaddress]{revtex4-1}
\usepackage{amsmath}
\usepackage{amsfonts}
\usepackage{amssymb}
\usepackage{latexsym}
\usepackage{enumerate}
\usepackage{color,xcolor}
\usepackage{graphicx}
\usepackage{bm}
\usepackage[english]{babel}
\usepackage{hyperref}
\usepackage{times}
\usepackage{comment}

\begin{document}

\title{The L\'{e}vy wave packet and fractional uncertainty relations}
\author{Zi-Xuan Li}
\author{Xiao-Dong Li}
\author{Towe Wang}
\email[Electronic address: ]{twang@phy.ecnu.edu.cn}
\affiliation{School of Physics, East China Normal University, Shanghai 200241, China\\ \vspace{0.2cm}}
\date{\today\\ \vspace{1cm}}
\begin{abstract}
This work revisits the debate on fractional uncertainty relations that arose a decade ago and tries to settle it. After a careful study of the position-momentum uncertainty product of the L\'{e}vy wave packet, a novel lower bound is proposed for uncertainty relations in fractional quantum mechanics. This bound is nearly linear in the L\'{e}vy index and is satisfied by exact, calculable solutions available in the literature.
\end{abstract}


\maketitle




\section{Introduction}\label{sect-intro}
As one of the major physics achievements of the last century, quantum mechanics has been confirmed by a large body of experiments. However, there are still various attempts to modify quantum mechanics, each motivated by different physical and mathematical considerations \cite{1990PhLB..234..276K,1992PhLB..293...37E,1998PhRvL..80.5243B,2000PhLA..268..298L,2022PhRvD.105e5002K}. In Ref. \cite{1990PhLB..234..276K}, Konishi et. al. established connections between string theory and observable quantum gravity effects through the generalized uncertainty principle. Later, Ellis and collaborators proposed that string theory fundamentally modifies quantum mechanics via quantum-gravitational friction from space-time foam \cite{1992PhLB..293...37E}. A distinctly different direction emerged when Bender and Boettcher demonstrated that non-Hermitian Hamiltonians with PT symmetry can exhibit entirely real spectra, revolutionizing the understanding of consistent quantum theories beyond the self-adjointness requirement \cite{1998PhRvL..80.5243B}. In Ref. \cite{2000PhLA..268..298L}, Laskin developed fractional quantum mechanics through L\'{e}vy path integrals, generalizing the Schr\"{o}dinger equation using fractional derivatives. Most recently, in Ref. \cite{2022PhRvD.105e5002K} Kaplan and Rajendran revisited nonlinear modifications of quantum mechanics while preserving causality. Among them, Laskin's fractional quantum mechanics is the focus of the present work.

The standard Schr\"{o}dinger equation in quantum mechanics, introduced one hundred years ago, are based on integer-order derivatives with respect to time and spatial coordinates. Adopting the space-fractional derivative, Laskin proposed a generalization of standard quantum mechanics, which is dubbed the fractional quantum mechanics and encapsulated by the so-called fractional Schr\"{o}dinger equation \cite{2000PhLA..268..298L,2000Chaos..10..780L,2000PhRvE..62.3135L,2002PhRvE..66e6108L}. In its Hamiltonian, the kinetic term is proportional to \(|p|^\alpha\), where \(1<\alpha\le 2\) is the L\'{e}vy index. The \(\alpha=2\) case recovers standard quantum mechanics, while the boundary \(\alpha=1\) is an unattainable limit that lies outside the domain of fractional quantum mechanics.


Since fractional-order differential equations are more challenging to solve analytically than their integer-order counterparts, understandably exact solutions to the fractional Schr\"{o}dinger equation have been found only in a limited number of cases. Early work established solutions for the free particle, infinite square well, and finite square barrier \cite{2000Chaos..10..780L,2006JMP....47h2104G}, alongside linear, delta-well, and Coulomb potentials \cite{2007JMP....48g2105D}. Subsequent studies further expanded this catalogue to include periodic potentials and the finite square well \cite{2008JMP....49e2105D}, and investigated single and double delta potentials \cite{2008JMP....49e2105D,2010JMP....51l3517D,2011JPhA...44r5303C}. More recent work has reported exact solutions for the infinite potential wall \cite{2021JMP....62c2104D}.

The uncertainty relation, formulated shortly after the Schr\"{o}dinger equation, played a central role in the development and interpretation of quantum mechanics. It remains a key tool for testing deviations from standard quantum mechanics. In the context of fractional quantum mechanics, Laskin introduced a generalized uncertainty relation, using the one-dimensional L\'{e}vy wave packet as a starting point~\cite{2000PhRvE..62.3135L}. A decade ago, there was a debate on the fractional uncertainty relation in fractional quantum mechanics, as presented in Refs.~\cite{2016PhRvE..93f6103W,2016PhRvE..93f6104L}. In Ref.~\cite{2016PhRvE..93f6103W}, it was shown that the fractional uncertainty relation proposed in Ref.~\cite{2000PhRvE..62.3135L} is inconsistent with a direct calculation using the one-dimensional L\'{e}vy wave packet for L\'{e}vy index \(\alpha = 1\). Ref.~\cite{2016PhRvE..93f6104L} defended the original fractional uncertainty relation of Ref.~\cite{2000PhRvE..62.3135L}.

The purpose of the present work is to settle the debate by preserving the spirit of Laskin's proposal while rectifying the calculations. By examining the position-momentum uncertainty product of the fractional L\'{e}vy wave packet analytically near the boundary \(\alpha=1\), we confirm that the lower bound in the original fractional uncertainty relation is violated. We then perform analytical calculations near the other boundary \(\alpha=2\) and numerical computations over the full parameter range. On this basis, we propose a novel bound for the fractional uncertainty relation and examine it with exact solutions available in the literature.

The remainder of this paper is organized as follows. Sec.~\ref{sect-rev} revisits Laskin's fractional uncertainty relation \cite{2000PhRvE..62.3135L} and identifies its underlying assumptions. Sec.~\ref{sect-ulev} examines the fractional uncertainty relation in the limit \(\alpha \to 1^+\), finding that the limit is smooth but the bound proposed by Laskin requires modification. A revised bound is then presented in Sec.~\ref{sect-mod}, where, after studying the position-momentum uncertainty product of the fractional L\'{e}vy wave packet both analytically and numerically, we put forward an approximate lower bound for fractional uncertainty relations. To examine this new bound, in Sec.~\ref{sect-test} we compute the position-momentum uncertainty product for two exact solutions of the fractional Schr\"{o}dinger equation corresponding to specific potentials: the infinite square potential well (Sec.~\ref{subsect-infwell}) and the delta potential well (Sec.~\ref{subsect-delwell}). Both solutions are shown to satisfy the new bound. Finally, Sec.~\ref{sect-disc} concludes the paper with a brief discussion.

Throughout this paper, we will repeatedly use the differentiation formulas
\begin{align}
\frac{d}{dz}a^z = a^z \ln a, \qquad \frac{d\Gamma(z)}{dz}\Big|_{z=1} = -\gamma,
\end{align}
where \(\Gamma(z)\) is the Gamma function and \(\gamma\) is the Euler--Mascheroni constant.

\section{Fractional uncertainty relation revisited}\label{sect-rev}
In standard quantum mechanics, the Gaussian wave packet is the minimum-uncertainty state that saturates Heisenberg's uncertainty relation. Naturally, its generalization in fractional quantum mechanics, the L\'{e}vy wave packet, is expected to play the analogous role in formulating and testing the fractional uncertainty relation. This possibility has been explored by Laskin in Ref.~\cite{2000PhRvE..62.3135L}. For a one-dimensional fractional L\'{e}vy wave packet, Laskin proved in Ref.~\cite{2000PhRvE..62.3135L} that the \(\mu\)-roots of the mean-\(\mu\) deviations of position and momentum are given respectively by
\begin{align}
\label{madx}\langle |\Delta x|^{\mu} \rangle^{1/\mu}_L &= \frac{l}{2^{1/\mu}} \mathcal{N}^{1/\mu}(\alpha, \mu, \nu; \tau, \eta_0),\\
\label{madp}\langle |\Delta p|^{\mu} \rangle^{1/\mu}_L &= \frac{\hbar}{l}\left[\frac{\Gamma\left(\frac{\mu+1}{\nu}\right)}{\Gamma\left(\frac{1}{\nu}\right)}\right]^{1/\mu},
\end{align}
where
\begin{align}\label{N}
\nonumber \mathcal{N}(\alpha, \mu, \nu; \tau, \eta_0) &= \frac{2^{1/\nu} \nu}{4\pi \Gamma\left(\frac{1}{\nu}\right)} \int_{-\infty}^{\infty} d\zeta \, |\zeta|^{\mu} \int_{-\infty}^{\infty} d\eta\\
\nonumber &\quad \times \int_{-\infty}^{\infty} d\eta' \exp\Bigl[i(\eta - \eta') \left(\zeta + \alpha \tau \eta_0^{\alpha-1}\right)\Bigr]\\
\nonumber &\quad \times \exp\Bigl[-i\tau\left(|\eta|^{\alpha} - |\eta'|^{\alpha}\right) - |\eta-\eta_0|^{\nu}\\
&\quad - |\eta'-\eta_0|^{\nu}\Bigr].
\end{align}
While the order of integration over \(\eta\) and \(\eta'\) can be interchanged, the integration over \(\zeta\) should be carried out last. The parameters obey \(\mu < \nu \leq \alpha\) with the L\'{e}vy index \(\alpha \in (1,2)\), in accordance with the properties of the L\'{e}vy \(\alpha\)-stable probability distribution. Furthermore,
\begin{align}\label{dxdp}
\nonumber &\Delta x = x - \langle x \rangle, \quad \Delta p = p - \langle p \rangle,\\
&\eta_0 = \frac{p_0 l}{2^{1/\nu} \hbar}, \quad \tau = \frac{D_{\alpha} t}{\hbar} \left( \frac{2^{1/\nu} \hbar}{l} \right)^\alpha.
\end{align}

On the basis of this result, Laskin claimed that
\begin{equation}\label{ass3}
\mathcal{N}(\alpha, \mu, \nu; \tau, \eta_0) > 1
\end{equation}
and \(\Gamma(\mu + 1/\nu)/\Gamma(1/\nu) \approx 1/\nu\), and proposed the fractional generalization of the well-known Heisenberg uncertainty relation,
\begin{equation}\label{fur}
\langle |\Delta x|^{\mu} \rangle^{1/\mu} \langle |\Delta p|^{\mu} \rangle^{1/\mu} > \frac{\hbar}{(2\alpha)^{1/\mu}}, \quad \mu < \alpha, \quad 1 < \alpha \leq 2,
\end{equation}
as one of the basic equations of fractional quantum mechanics \cite{2000PhRvE..62.3135L}. For the special case \(\alpha = 2\) one can set \(\mu = 2\), which is not at odds with the properties of the L\'{e}vy \(\alpha\)-stable probability distribution.

In the standard quantum mechanical limit (\(\alpha = 2,\ \mu = 2\)), the relation \eqref{fur} reduces to the Heisenberg uncertainty principle
\begin{equation}\label{fur2}
\langle |\Delta x|^2 \rangle^{1/2} \langle |\Delta p|^2 \rangle^{1/2} > \frac{\hbar}{2}.
\end{equation}
In the other limit (\(\alpha = 1,\ \mu = 1\)), it tends to
\begin{equation}\label{fur1}
\langle |\Delta x| \rangle \langle |\Delta p| \rangle > \frac{\hbar}{2}.
\end{equation}
Although the lower bounds on the right-hand sides of \eqref{fur2} and \eqref{fur1} are identical, the left-hand sides involve different measures of uncertainty: the standard deviation in the first case and the mean absolute deviation in the second.

Taking a closer look at the procedures above, we can find that the following assumptions have been made when reproducing the standard Heisenberg uncertainty relation from the L\'{e}vy wave packet:
\begin{enumerate}[(1)]
\item With \(\mu < \alpha\), the position-momentum uncertainty product \(\langle |\Delta x|^{\mu} \rangle^{1/\mu} \langle |\Delta p|^{\mu} \rangle^{1/\mu}\) has a lower bound.
\item The lower bound is less than or equal to \(\langle |\Delta x|^{\mu} \rangle^{1/\mu}_L \langle |\Delta p|^{\mu} \rangle^{1/\mu}_L\).
\item The inequality \eqref{ass3} is assumed to be valid.
\item When writing down \eqref{fur}, one imposes \(\nu = \alpha\).
\item The Heisenberg uncertainty relation is the \(\mu \to 2^-\) limit of the fractional uncertainty relation. Since the parameters must satisfy \(\mu < \alpha \leq 2\) (as \(\nu = \alpha\)), this limit can be approached only if \(\alpha \to 2^-\).
\end{enumerate}
In addition, two more assumptions, though often overlooked, must be made for consistency:
\begin{enumerate}[(1)]\setcounter{enumi}{5}
\item Since the lower bound of the standard Heisenberg uncertainty relation is attainable, the strict inequality ``\(>\)" in the fractional uncertainty relation \eqref{fur} should be replaced by ``\(\geq\)".
\item It is well known that the position uncertainty of a free Gaussian wave packet grows with time in standard quantum mechanics. It is therefore more reasonable to evaluate \(\langle |\Delta x|^{\mu} \rangle^{1/\mu}_L \langle |\Delta p|^{\mu} \rangle^{1/\mu}_L\) in assumption (2) at the initial time \(t=0\), i.e., \(\tau=0\).
\end{enumerate}

In Ref.~\cite{2016PhRvE..93f6103W}, Wei suggested replacing the condition \(\mu < \alpha\) in assumption (1) with \(\mu = 2\) for arbitrary \(\alpha\), stating that ``the uncertainty relation is independent of wave equations''. This suggestion was refuted by Laskin in Ref.~\cite{2016PhRvE..93f6104L} on the grounds that the L\'{e}vy \(\alpha\)-stable probability distribution does not allow \(\mu \geq \alpha\) for \(\alpha < 2\).

Moreover, Wei proved that \(\langle |\Delta x| \rangle_L \langle |\Delta p| \rangle_L = \hbar / \pi\), and consequently, the extrapolation of \eqref{fur} to \(\alpha = 1,\ \mu = 1\) contradicts assumption (2). This issue was not directly resolved in Ref.~\cite{2016PhRvE..93f6104L}. As we will demonstrate in this paper, the root of the problem lies in assumption (3). Indeed, the claimed inequality~\eqref{ass3} is not valid. Correcting this inequality leads to an improved version of Laskin's fractional uncertainty relation~\eqref{fur} with a modified lower bound, which constitutes one of the main results of this paper.

While the fractional uncertainty relation is examined carefully at the limiting point \(\alpha = 1,\ \mu = 1\) in Ref.~\cite{2016PhRvE..93f6103W}, a subtlety must be addressed: fractional quantum mechanics is rigorously defined only for the L\'{e}vy index \(\alpha\) in the range \(1 < \alpha \le 2\). Consequently, to properly examine the uncertainty relation \eqref{fur} within this framework without relying on an extrapolation to the boundary, it is necessary to evaluate the position-momentum uncertainty of the L\'{e}vy wave packet for \(\alpha > 1\). We therefore perform analytical calculations near \(\alpha = 1\) and \(\mu = 1\) in the next section. Numerical computations over the full parameter range \(1 < \alpha \le 2\) are deferred to Sec.~\ref{sect-mod}, where we restrict ourselves to \(\mu = \alpha^-\).

\section{Uncertainty of the L\'{e}vy wave packet near \(\alpha = 1\)}\label{sect-ulev}

In the present section, we perform an asymptotic expansion of \(\mathcal{N}(\alpha,\mu,\alpha;\tau,\eta_0)\) for \(\alpha = 1+\delta+\varepsilon,\ \mu = 1+\delta\), where \(\delta\) and \(\varepsilon\) are infinitesimal positive numbers. In this expansion, we keep only terms up to first order in \(\delta\) and \(\varepsilon\); quadratic and higher-order contributions are neglected. Subsequently, the position-momentum uncertainty of the L\'{e}vy wave packet is examined carefully in the same limit.

\subsection{Asymptotic expansion of \(\mathcal{N}(1+\delta+\varepsilon,1+\delta,1+\delta+\varepsilon;0,\eta_0)\)}
We start from the expression \eqref{N} with \(\tau=0\) and \(\nu=\alpha\):
\begin{align}\label{Nass5}
\nonumber \mathcal{N}(\alpha,\mu,\alpha;0,\eta_0)&=\frac{2^{1/\alpha}\alpha}{4\pi\,\Gamma(1/\alpha)}
\int_{-\infty}^{\infty} d\zeta\,|\zeta|^{\mu}\, \\
&\quad \times \biggl|\int_{-\infty}^{\infty} du\, \exp\Bigl(iu\zeta-|u|^{\alpha}\Bigr)\biggr|^2,
\end{align}
where we have made the change of variable \(u = \eta - \eta_0\) with \(\eta_0\) an arbitrary constant. Now the expression is actually independent of \(\eta_0\). We then introduce the infinitesimal positive numbers \(\delta\) and \(\varepsilon\) and set
\begin{equation}
\alpha = 1+\Delta,\qquad \mu = 1+\delta,\qquad \Delta=\delta+\varepsilon,
\end{equation}
with \(\delta,\varepsilon \to 0^+\). This parametrization ensures that \(\alpha\) and \(\mu\) approach \(1\) from above, and that \(\alpha - \mu = \varepsilon > 0\).

In the remainder of this subsection, we expand both the prefactor and the \(\zeta\)-integral appearing in Eq.~\eqref{Nass5} as power series in the small parameters \(\delta\) and \(\Delta\), keeping all terms up to first order. Multiplying these expansions yields a linear asymptotic expression for \(\mathcal{N}\) in terms of \(\delta\) and \(\varepsilon\).

The prefactor in Eq.~\eqref{Nass5} depends on \(\alpha\) solely through \(\Delta = \alpha-1\), and its expansion is a power series in \(\Delta\) with no other small parameters. A straightforward expansion gives
\begin{equation}\label{pref}
\frac{2^{1/\alpha}\alpha}{4\pi\,\Gamma(1/\alpha)} = \frac{1}{2\pi}\bigl[1 + a_1\Delta + O(\Delta^2)\bigr],
\end{equation}
where \(a_1 = 1 - \ln2 - \gamma\), and we have used the expansion \(\Gamma(1-z)=1+\gamma z+O(z^2)\) with \(z=1-1/\alpha = \Delta + O(\Delta^2)\).

In analogy with the prefactor, the inner \(u\)-integral
\begin{equation}\label{I}
I(\zeta)=\int_{-\infty}^{\infty} du\, \exp\Bigl(iu\zeta-|u|^{\alpha}\Bigr)
\end{equation}
depends solely on the small parameter \(\Delta = \alpha-1\), apart from the variable \(\zeta\). Setting \(\alpha = 1+\Delta\), we expand the exponent in the integrand of \(I(\zeta)\) systematically to first order in \(\Delta\). The phase factor \(iu\zeta\) is independent of \(\Delta\), while the nonlinear term is expanded as
\begin{equation}
|u|^{\alpha}=|u|+\Delta|u|\ln|u|+O(\Delta^2).
\end{equation}
Substituting this expansion into the exponent and retaining only terms linear in \(\Delta\) gives
\begin{equation}\label{Iexp}
I(\zeta)=I_0(\zeta)+\Delta I_1(\zeta)+O(\Delta^2),
\end{equation}
where the coefficient functions are defined by
\begin{align}\label{I01}
\nonumber I_0(\zeta) &= \int_{-\infty}^{\infty} du\, e^{iu\zeta-|u|}, \\
I_1(\zeta) &= - \int_{-\infty}^{\infty} du\, e^{iu\zeta-|u|}\, |u|\ln|u|.
\end{align}
Consequently, the squared modulus of \(I(\zeta)\) is
\begin{equation}
|I(\zeta)|^2 = |I_0(\zeta)|^2 + 2\Delta\,\operatorname{Re}\bigl(I_0^*(\zeta)I_1(\zeta)\bigr) + O(\Delta^2).
\end{equation}

The evaluation of the \(\zeta\)-integral in Eq.~\eqref{Nass5} proceeds by expanding the factor \(|\zeta|^{\mu}\). Since \(\mu = 1+\delta\), we have
\begin{equation}
|\zeta|^{\mu}=|\zeta|\,e^{\delta\ln|\zeta|}=|\zeta|\bigl[1+\delta\ln|\zeta|+O(\delta^2)\bigr].
\end{equation}
For compactness, we introduce the following auxiliary integrals, all evaluated at the unperturbed point \(\Delta=\delta=0\):
\begin{align}\label{Jlin}
\nonumber J_{00}&=\int_{-\infty}^{\infty} d\zeta\,|\zeta|\,|I_0|^2, \\
\nonumber J_{01}&=\int_{-\infty}^{\infty} d\zeta\,|\zeta|\ln|\zeta|\,|I_0|^2, \\
J_{10}&=2\int_{-\infty}^{\infty} d\zeta\,|\zeta|\,\operatorname{Re}(I_0^*I_1).
\end{align}
These integrals are \(\eta_0\)-independent constants. Substituting the expansions of \(|\zeta|^{\mu}\) and \(|I(\zeta)|^2\) and using the definitions above, the \(\zeta\)-integral can be expanded up to first order in \(\delta\) and \(\Delta\) as
\begin{equation}\label{dzeta}
\int_{-\infty}^{\infty} d\zeta\,|\zeta|^{\mu}|I(\zeta)|^2 = J_{00} + \delta J_{01} + \Delta J_{10} + O(\delta^2,\varepsilon^2,\delta\varepsilon)
\end{equation}
with \(\Delta = \delta+\varepsilon\).

The full expression for \(\mathcal{N}\) is the product of the prefactor expansion \eqref{pref} and the \(\zeta\)-integral \eqref{dzeta}. Collecting all terms up to first order in the small parameters \(\delta\) and \(\varepsilon\) (recall \(\Delta = \delta+\varepsilon\)), we obtain
\begin{align}
\nonumber \mathcal{N}(\alpha, \mu, \eta_0) &= \frac{1}{2\pi}\Bigl( J_{00} + \delta J_{01} + \Delta J_{10} + a_1\Delta J_{00} \Bigr) \\
&\qquad + O(\delta^2,\varepsilon^2,\delta\varepsilon),
\end{align}
where we have suppressed the arguments \(\nu\) and \(\tau\) by setting them to \(\alpha\) and \(0\), respectively. Substituting \(\Delta = \delta + \varepsilon\) and grouping terms of the same order in \(\delta\) and \(\varepsilon\) yields
\begin{align}\label{Nexp}
\nonumber \mathcal{N}(\alpha, \mu, \eta_0) &= \frac{J_{00}}{2\pi} + \frac{1}{2\pi}\Bigl[\bigl(J_{01}+J_{10}+a_1J_{00}\bigr)\,\delta \\
&\qquad + \bigl(J_{10}+a_1J_{00}\bigr)\,\varepsilon\Bigr] + O(\delta^2,\varepsilon^2,\delta\varepsilon).
\end{align}
The constant term \(J_{00}/(2\pi)\) equals the value of \(\mathcal{N}\) at \(\alpha=\mu=1\), namely \(\mathcal{N}(1,1,1;0,\eta_0)\). For \(\tau=0\), this constant is independent of \(\eta_0\), as noted above. The linear terms provide the leading corrections for small \(\delta\) and \(\varepsilon\).

\subsection{Position-momentum uncertainty of the L\'{e}vy wave packet}\label{subsect-ur0}
As detailed in Appendix~\ref{app-coef}, the coefficient functions~\eqref{I01} can be evaluated explicitly as
\begin{align}\label{Iin}
\nonumber I_0(\zeta) &= \int_{-\infty}^{\infty} du\; e^{iu\zeta - |u|} = \frac{2}{1+\zeta^2},\\
\nonumber I_1(\zeta) &= - \int_{-\infty}^{\infty} du\; e^{iu\zeta - |u|}\, |u|\ln|u|\\
&= \frac{1-\zeta^2}{(1+\zeta^2)^2}\bigl[\ln(1+\zeta^2)-2(1-\gamma)\bigr] + \frac{4\zeta\arctan\zeta}{(1+\zeta^2)^2}.
\end{align}

Having obtained explicit expressions for \(I_0\) and \(I_1\), we now evaluate the relevant auxiliary integrals \eqref{Jlin}. A direct calculation gives
\begin{equation}
J_{00}=\int_{-\infty}^{\infty} d\zeta\,|\zeta|\,\frac{4}{(1+\zeta^2)^2}=4.
\end{equation}
The integral
\begin{equation}
J_{01} = 8\int_0^{\infty}\frac{\zeta\ln\zeta}{(1+\zeta^2)^2}d\zeta
\end{equation}
vanishes because the change of variable \(\zeta \to 1/\zeta\) maps the integral to its negative, giving \(J_{01} = -J_{01}\) and therefore \(J_{01}=0\).

The value of \(J_{10}\) follows from a direct integration
\begin{equation}
J_{10}=4\int_0^{\infty}\zeta\,I_0(\zeta)I_1(\zeta)\,d\zeta,
\end{equation}
in which
\begin{align}
\nonumber \zeta\,I_0(\zeta)I_1(\zeta) &= \frac{2\zeta(1-\zeta^2)}{(1+\zeta^2)^3}\bigl[\ln(1+\zeta^2)-2(1-\gamma)\bigr] \\
\quad &+ \frac{8\zeta^2\arctan\zeta}{(1+\zeta^2)^3}.
\end{align}
The first term is elementary: the substitution \(s = 1+\zeta^2\) transforms it into a rational integral that can be evaluated straightforwardly. The second term involves the standard integral
\begin{equation}
\int_0^{\infty}\frac{\zeta^2\arctan\zeta}{(1+\zeta^2)^3}\,d\zeta,
\end{equation}
which is often evaluated by setting \(\zeta = \tan\theta\) and using trigonometric identities. Carrying out both integrals and simplifying the result yields
\begin{equation}
J_{10}= \frac{\pi^2}{2}-2.
\end{equation}
The detailed steps are routine and can be found in integral tables; they are omitted here for brevity.

Substituting the values \(a_1 = 1 - \ln2 - \gamma,\ J_{00}=4,\ J_{01}=0\) and \(J_{10} = \pi^2/2 - 2\) into the linear expansion \eqref{Nexp} yields
\begin{align}\label{Nin}
\nonumber \mathcal{N}(\alpha, \mu, \eta_0) &= \frac{2}{\pi} + \frac{1}{2\pi}\Bigl(\frac{\pi^2}{2} + 2 - 4\ln2 - 4\gamma\Bigr)(\delta+\varepsilon) \\
\quad &+ \mathcal{O}(\delta^2,\varepsilon^2,\delta\varepsilon).
\end{align}
Thus, for initial states with \(t = 0\), that is \(\tau = 0\), the first-order term in \(\mathcal{N}\) depends linearly on \(\delta+\varepsilon\) near \(\alpha = \mu = 1\).

We now expand the full uncertainty product for the initial L\'{e}vy wave packet
\begin{equation}\label{Lup}
\langle |\Delta x|^{\mu} \rangle^{1/\mu}_L \langle |\Delta p|^{\mu} \rangle^{1/\mu}_L
= \hbar \left[ \frac{\Gamma\!\left(\frac{\mu+1}{\alpha}\right)}{2\,\Gamma\!\left(\frac{1}{\alpha}\right)} \,\mathcal{N} \right]^{1/\mu}
\end{equation}
to first order in the small parameters \(\delta\) and \(\varepsilon\) with \(\alpha = 1+\delta+\varepsilon\) and \(\mu = 1+\delta\).  Using Eq.~\eqref{Nin} for \(\mathcal{N}\) together with the expansion
\begin{equation}
\frac{\Gamma\!\left(\frac{\mu+1}{\alpha}\right)}{\Gamma\!\left(\frac{1}{\alpha}\right)} = 1 - \delta - (2-\gamma)\varepsilon + O(\delta^2,\varepsilon^2,\delta\varepsilon)
\end{equation}
and retaining only first-order terms, we obtain the linear approximation for the bracket
\begin{align}
\nonumber \frac{\Gamma\!\left(\frac{\mu+1}{\alpha}\right)}{2\,\Gamma\!\left(\frac{1}{\alpha}\right)} \,\mathcal{N}
&= \frac{1}{\pi}\Bigl[1 + \Bigl(\frac{\pi^2}{8} - \frac12 - \ln2 - \gamma\Bigr)\delta \\
&\quad + \Bigl(\frac{\pi^2}{8} - \frac32 - \ln2\Bigr)\varepsilon\Bigr] + O(\delta^2,\varepsilon^2,\delta\varepsilon).
\end{align}
Consequently the linear expansion of the uncertainty product itself is found to be
\begin{align}\label{Lup1}
\nonumber \langle |\Delta x|^{\mu} \rangle^{1/\mu}_L \langle |\Delta p|^{\mu} \rangle^{1/\mu}_L
&= \frac{\hbar}{\pi}\Bigl[1 + \Bigl(\frac{\pi^2}{8} - \frac12 + \ln\frac{\pi}{2} - \gamma\Bigr)\delta \\
\nonumber &\quad + \Bigl(\frac{\pi^2}{8} - \frac32 - \ln2\Bigr)\varepsilon\Bigr] \\
\nonumber &\quad + O(\delta^2,\varepsilon^2,\delta\varepsilon) \\
&\approx \frac{\hbar}{\pi}(1 + 0.61 \delta - 0.96\varepsilon).
\end{align}
Hence the uncertainty product near \(\alpha = \mu = 1\) has independent linear dependencies on \(\delta\) and \(\varepsilon\).

This result extends Equation (12) in Ref.~\cite{2016PhRvE..93f6103W} from \(\alpha=1\) to \(\alpha>1\), a region where fractional quantum mechanics is rigorously defined. For comparison, the lower bound of~\eqref{fur} is
\begin{equation}
\frac{\hbar}{(2\alpha)^{1/\mu}} = \frac{\hbar}{2}\Bigl[1 + (\ln2 - 1)\delta - \varepsilon\Bigr] + \mathcal{O}(\delta^2,\varepsilon^2,\delta\varepsilon),
\end{equation}
with \(\alpha = 1+\delta+\varepsilon,\ \mu = 1+\delta\). This expression is at odds with Eq.~\eqref{Lup1}, indicating that the estimate in \eqref{ass3} is violated. As a result, the lower bound in~\eqref{fur} does not actually represent the true position-momentum uncertainty product of the L\'{e}vy wave packet.

\section{Fractional uncertainty relation revised}\label{sect-mod}
\subsection{Uncertainties of the L\'{e}vy wave packet with quadratic corrections near \(\alpha = 1, 2\)}\label{subsect-ana}
The calculations in the previous section can be extended straightforwardly to the second order in \(\delta\) and \(\varepsilon\), as well as to the opposite boundary \(\alpha = 2\). Since the full details are too lengthy to present here, we merely summarize the final results. In both cases, we set \(\nu = \alpha\) and \(\tau = 0\).

Near the boundary \(\alpha = 1\), after some algebra, we obtain the analytic expressions for the second-order terms in Eq.~\eqref{Lup1},
\begin{align}\label{Lup2}
\nonumber &\frac{\hbar}{\pi} \biggl\{ \Bigl[ \Bigl(\frac{\pi^2}{8} - \frac{3}{2}\Bigr)\Bigl(\ln\frac{\pi}{2} - \gamma - \frac{16}{9}\Bigr) + \frac12\Bigl(\ln\frac{\pi}{2} - \gamma\Bigr)^2 + \ln2 \\
\nonumber &+ \frac{7}{8}\zeta(3) - \frac{7}{3} \Bigr] \delta^2 + \Bigl[ \frac12\ln^2 2 + \Bigl(\frac{5}{2} - \frac{\pi^2}{8}\Bigr)\ln2 -\frac{\pi^2}{72} + \frac{5}{6} \Bigr] \varepsilon^2 \\
\nonumber &+ \Bigl[ \Bigl(\frac{\pi^2}{8} - \frac{3}{2}\Bigr)\Bigl(\ln\frac{\pi}{4} - \gamma - \frac{26}{9}\Bigr) + \Bigl(\gamma + 2 - \ln\frac{\pi}{2}\Bigr)\ln2 \\
&+ \frac{7}{8}\zeta(3) - \frac{13}{6} \Bigr]\delta\varepsilon
\end{align}
with the infinitesimal positive parameters \(\delta = \mu - 1\) and \(\varepsilon = \alpha - \mu\). Here \(\zeta(3)\) is the Riemann zeta function evaluated at 3, also known as Ap\'{e}ry's constant.

Near the other boundary \(\alpha = 2\), with the notation \(\delta = 2 - \alpha\) and \(\varepsilon = \alpha - \mu\), we find
\begin{align}\label{Hup}
\nonumber \langle |\Delta x|^{\mu} \rangle^{1/\mu}_L \langle |\Delta p|^{\mu} \rangle^{1/\mu}_L
&= \frac{\hbar}{2}\Bigl[1 + \Bigl(\frac{\gamma}{2} -1 + \frac12 \ln2\Bigr)\delta \\
\nonumber &\quad + \Bigl(\frac{\gamma}{2} -1 + \frac14 \ln2\Bigr)\varepsilon\Bigr] \\
\nonumber &\quad + O(\delta^2,\varepsilon^2,\delta\varepsilon) \\
&\approx \frac{\hbar}{2}(1 - 0.36\delta - 0.54\varepsilon).
\end{align}

For later use, we specialize to \(\mu = \alpha\) and introduce the notation
\begin{equation}
\Pi = \langle |\Delta x|^{\alpha} \rangle^{1/\alpha} \langle |\Delta p|^{\alpha} \rangle^{1/\alpha},
\end{equation}
and combine the two limiting behaviors as
\begin{widetext}
\begin{equation}\label{LHup}
\Pi_L =
\begin{cases}
\begin{aligned}
&\frac{\hbar}{\pi}\Biggl\{ 1 + \Bigl(\frac{\pi^2}{8} - \frac12 + \ln\frac{\pi}{2} - \gamma\Bigr)(\alpha-1) \\
&\quad + \Bigl[ \Bigl(\frac{\pi^2}{8} - \frac{3}{2}\Bigr)\Bigl(\ln\frac{\pi}{2} - \gamma - \frac{16}{9}\Bigr) + \frac12\Bigl(\ln\frac{\pi}{2} - \gamma\Bigr)^2 + \ln2 + \frac{7}{8}\zeta(3) - \frac{7}{3} \Bigr] (\alpha-1)^2 \Biggr\},
\end{aligned} & \alpha \to 1,\\[6pt]
\begin{aligned}
&\frac{\hbar}{2}\Bigl[1 + \Bigl(\frac{\gamma}{2} -1 + \frac12 \ln2\Bigr)(2-\alpha) \Bigr],
\end{aligned} & \alpha \to 2.
\end{cases}
\end{equation}
\end{widetext}

\subsection{Uncertainty of the L\'{e}vy wave packet for \(1 < \alpha \le 2\)}\label{subsect-num}
\begin{figure}[htbp]
    \centering
    \includegraphics[width=0.4\textwidth]{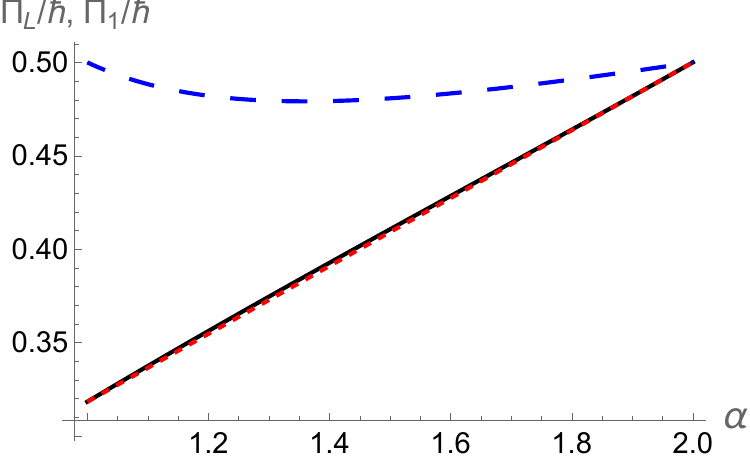}\\
    \caption{The position-momentum uncertainty product of the L\'{e}vy wave packet~\eqref{Lup} with \(\mu=\alpha^-\) (\(\Pi_L\), black solid line), the linear lower bound from the revised fractional uncertainty relation~\eqref{newfur1} (\(\Pi_1\), red dashed line), and Laskin's original bound~\eqref{fur} with \(\mu=\alpha^-\) (blue long-dashed curve). The refined lower bound~\eqref{newfur3} (\(\Pi_3\)) is not plotted here, because it is visually identical to the black solid line within the resolution of the figure.}\label{fig:Lup}
\end{figure}

While the asymptotic behavior of Eq.~\eqref{Lup} near the boundaries can be derived analytically, a closed-form solution over the entire interval \(1 < \alpha \le 2\) is unattainable, and one must resort to numerical simulations. The numerical result is depicted by the black solid curve in Fig.~\ref{fig:Lup}. Despite the intricate analytic structure of Eq.~\eqref{Lup}, the simulated behavior is remarkably simple: \(\Pi_L\) is almost linear in \(\alpha\) across the whole range.

In the same figure, we also plot a linear function (red dashed line) connecting the two endpoints of the black solid curve, namely \(\Pi(1)=\hbar/\pi\) and \(\Pi(2)=\hbar/2\). This line almost coincides with the black solid curve, and its explicit form is given by Eq.~\eqref{newfur1} in the next subsection. For comparison, we also plot Laskin's original bound, i.e., the right-hand side of~\eqref{fur} evaluated at \(\mu=\alpha^-\), shown as the blue long-dashed curve. This curve lies visibly above the black solid curve, confirming that the original bound is violated over the full parameter range.

\subsection{A revised fractional uncertainty relation}\label{sect-new}
It is now clear that assumption (3) in Sec.~\ref{sect-rev} is untenable and must be discarded. Guided by the numerical findings of the previous subsection, we replace \(\mu < \alpha\) with \(\mu = \alpha - 0^+\) in assumption (1), so as to keep the formulation both elegant and conservative. With these modifications, the fractional uncertainty relation can be revised as follows:

\emph{In Laskin's fractional quantum mechanics with L\'{e}vy index \(\alpha \in (1,2)\), the position-momentum uncertainty product satisfies the lower bound}
\begin{equation}\label{newfur1}
\lim_{\mu\rightarrow\alpha}\langle |\Delta x|^{\mu} \rangle^{1/\mu} \langle |\Delta p|^{\mu} \rangle^{1/\mu} \geq \frac{\hbar}{2\pi}\bigl[2 + (\pi-2)(\alpha-1)\bigr].
\end{equation}

\begin{figure}[htbp]
    \centering
    \includegraphics[width=0.4\textwidth]{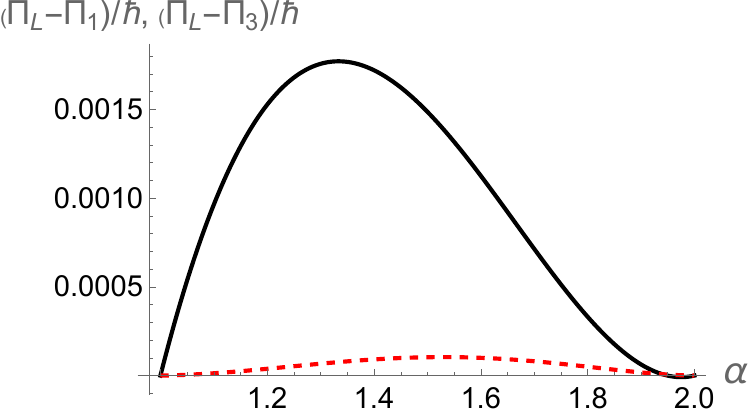}\\
    \caption{Residuals \(\Pi_L-\Pi_1\) (the black solid curve) and \(\Pi_L-\Pi_3\) (the red dashed curve), where \(\Pi_L\) is the L\'{e}vy wave packet uncertainty product~\eqref{Lup} with \(\mu=\alpha\), and \(\Pi_1,\, \Pi_3\) are the linear and refined lower bounds~\eqref{newfur1},~\eqref{newfur3}.}\label{fig:dup}
\end{figure}

Comparing this bound with Eq.~\eqref{LHup}, we find that the zeroth-order terms are in accordance with each other. At first glance, as illustrated in Fig.~\ref{fig:Lup}, the bound seems to agree very well with the L\'{e}vy wave packet uncertainty product~\eqref{Lup}. However, a closer look reveals that the first-order coefficients in~\eqref{Lup} and~\eqref{newfur1} satisfy the numerical ordering
\begin{equation}
\pi-2 < -\pi\Bigl(\frac{\gamma}{2} -1 + \frac12 \ln2\Bigr) < \Bigl(\frac{\pi^2}{8} - \frac12 + \ln\frac{\pi}{2} - \gamma\Bigr),
\end{equation}
which indicates that the bound~\eqref{newfur1} is violated near \(\alpha = 2\) by the L\'{e}vy wave packet. The small violation is highlighted in Fig.~\ref{fig:dup}, where the black solid curve shows the difference \(\Pi_L - \Pi_1\), i.e., the numerically obtained value of~\eqref{Lup} (with \(\mu=\alpha\)) minus the right-hand side of~\eqref{newfur3}.

In Fig.~\ref{fig:dup}, the black solid curve resembles a segment of a cubic polynomial. In certain intervals, the difference becomes negative, signaling a breakdown of the bound \(\Pi_1\). To remedy this, we fit the linear terms in Eq.~\eqref{LHup} with a cubic polynomial, yielding the refined lower bound
\begin{equation}\label{newfur3}
\frac{\hbar}{2\pi}\bigl[2 + (\pi-2)(\alpha-1) + 0.079(\alpha-1)(\alpha-2)(\alpha-1.94)\bigr]
\end{equation}
which will be denoted by \(\Pi_3\). Indeed, over the interval \(1.94 < \alpha < 2\), this bound is lower than the linear bound \(\Pi_1\), thereby remedying the violation near \(\alpha = 2\).  In addition, the cubic bound significantly reduces the theoretical estimation error. This is evident in Fig.~\ref{fig:dup}, where the residual \(\Pi_L-\Pi_3\) is far smaller than \(\Pi_L-\Pi_1\) yet remains positive, demonstrating that the refined bound provides a better estimate of the numerical result. Specifically, the maximum relative estimation error is reduced by a factor of \(40\), from \(0.4\%\) to \(0.01\%\).

\section{Testing against exact solutions}\label{sect-test}
\subsection{Conventions and methodology}\label{subsect-meth}
In one dimension, the Hamiltonian of fractional quantum mechanics takes the form
\begin{equation}
H = D_\alpha \left(-\hbar^2 \frac{d^2}{dx^2}\right)^{\alpha/2} + V(x),
\end{equation}
where \(V(x)\) is the potential. The stationary wavefunctions in coordinate and momentum spaces, \(\psi(x)\) and \(\phi(p)\), respectively, are related by
\begin{align}\label{Fourier}
\nonumber \psi(x) &= \frac{1}{2\pi\hbar} \int_{-\infty}^{\infty} dp \, e^{i p x/\hbar}\,\phi(p),\\
\phi(p) &= \int_{-\infty}^{\infty} dx \, e^{-i p x/\hbar}\,\psi(x).
\end{align}
They satisfy the normalization conditions \(\int_{-\infty}^{\infty} dx \, |\psi(x)|^2 = 1\) and \(\int_{-\infty}^{\infty} dp \, |\phi(p)|^2 = 2\pi\hbar\). The mean values and the mean-\(\mu\) absolute deviations of position and momentum are given by
\begin{align}\label{mean}
\nonumber &\langle x \rangle = \int_{-\infty}^{\infty} dx \, x \, |\psi(x)|^2 = \frac{i}{2\pi} \int_{-\infty}^{\infty} dp \, \phi^*(p)\phi'(p),\\
\nonumber &\langle p \rangle = \frac{1}{2\pi\hbar}\int_{-\infty}^{\infty} dp \, p \, |\phi(p)|^2 = -i\hbar \int_{-\infty}^{\infty} dx \, \psi^*(x)\psi'(x),\\
\nonumber &\langle |\Delta x|^{\mu} \rangle = \int_{-\infty}^{\infty} dx \, |x - \langle x \rangle|^{\mu} \, |\psi(x)|^2,\\
&\langle |\Delta p|^{\mu} \rangle = \frac{1}{2\pi\hbar}\int_{-\infty}^{\infty} dp \, |p - \langle p \rangle|^{\mu} \, |\phi(p)|^2.
\end{align}

If the probability densities are even, i.e., \(|\psi(-x)|^2=|\psi(x)|^2\) and \(|\phi(-p)|^2=|\phi(p)|^2\), then the mean values vanish, \(\langle x \rangle = \langle p \rangle = 0\), and the mean-\(\mu\) deviations reduce to absolute moments,
\begin{align}\label{meanab}
\nonumber \langle |\Delta x|^{\mu} \rangle
&= 2\int_{0}^{\infty} dx \, x^{\mu}\,|\psi(x)|^2,\\
\langle |\Delta p|^{\mu} \rangle
&= \frac{1}{\pi\hbar}\int_{0}^{\infty} dp \, p^{\mu}\,|\phi(p)|^2,
\end{align}
provided the integrals converge. For L\'{e}vy-stable wave packets, this requires \(\mu<\alpha\).

Having established our conventions and methodology, we are now ready to test the revised fractional uncertainty relation, which is restricted to \(\mu=\alpha^-\), against exact solutions of the fractional Schr\"{o}dinger equation available in the literature. We have reviewed a number of these solutions in Sec.~\ref{sect-intro}. Unfortunately, most solutions, such as the free particle, cannot be normalized to a finite value; the exceptions are the square potential well, the delta potential well, and the Coulomb potential problems. What is more, the solutions for the finite square potential well and the Coulomb potential problems are rather intricate and primitive, and we therefore postpone an exhaustive investigation of them to another work. In the remainder of this work, we examine the infinite square potential well in Sec.~\ref{subsect-infwell} and the delta potential well in Sec.~\ref{subsect-delwell}.

\subsection{Infinite square potential well}\label{subsect-infwell}
In fractional quantum mechanics, a particle confined in an infinite potential well
\begin{equation}
V(x) =
\begin{cases}
0, & |x| < a,\\
\infty, & |x| \ge a
\end{cases}
\end{equation}
can be described by the wavefunctions \cite{2000Chaos..10..780L,2006JMP....47h2104G}
\begin{equation}
\psi_n(x) =
\begin{cases}
\frac{1}{\sqrt{a}} \sin \frac{n\pi}{2a}(x+a), & |x| < a,\\
0, & |x| \ge a
\end{cases}
\end{equation}
with \(n=1,2,3,\ldots\). The wavefunctions are normalized and mutually orthogonal inside the well and vanish outside. Using Eq.~\eqref{Fourier}, one can directly obtain the momentum-space wavefunctions
\begin{equation}
\phi_n(p)=\frac{2n\sqrt{a}}{\pi}\,
\frac{e^{ipa/\hbar}-(-1)^n e^{-ipa/\hbar}}{n^2-\left[2pa/(\pi\hbar)\right]^2}.
\end{equation}

\begin{figure}[htbp]
    \centering
    \includegraphics[width=0.4\textwidth]{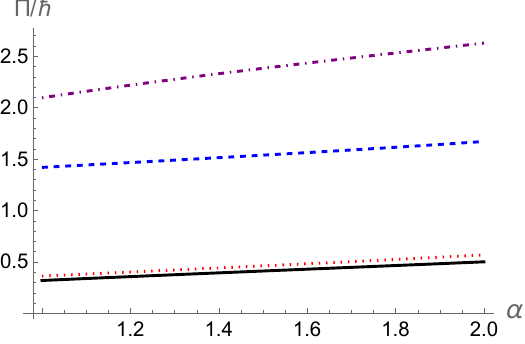}\\
    \caption{Position-momentum uncertainty product with \(\mu=\alpha^-\) for the three lowest energy states of the infinite square potential well (red dotted, blue dashed, and purple dash-dotted lines for \(n=1,2,3\), respectively) and for the L\'{e}vy wave packet~\eqref{Lup} (black solid line).}\label{fig:infwell}
\end{figure}

Because the probability densities \(|\psi_n(x)|^2\) and \(|\phi_n(p)|^2\) are even functions here, integrating \(x|\psi_n(x)|^2\) and \(p|\phi_n(p)|^2\) over the symmetric interval yields zero, and hence \(\langle x \rangle = \langle p \rangle = 0\). Consequently, the mean-\(\mu\) deviations of \(x\) and \(p\) reduce to the absolute moments~\eqref{meanab}. These integrals cannot be evaluated in closed form, so we perform numerical computations and present the results in Fig.~\ref{fig:infwell}. As can be seen, the position-momentum uncertainty products \(\Pi\) for the \(n=1,2,3\) states all lie above \(\Pi_L\) of the L\'{e}vy wave packet, thereby satisfying the revised uncertainty relation \(\Pi \ge \Pi_L\) at \(\mu=\alpha^-\).

\subsection{Delta potential well}\label{subsect-delwell}
In a delta potential well, \(V(x)=-V_0\delta(x)\) with \(V_0>0\), a particle has a single bound state with the unique energy level
\begin{equation}
E=-\left[\frac{V_0\,B(1/\alpha,\,1-1/\alpha)}{\pi\hbar\alpha D_\alpha^{1/\alpha}}\right]^{\alpha/(\alpha-1)},
\end{equation}
where \(B(\cdot,\cdot)\) denotes the Euler beta function, which can be evaluated as \(B\left(1/\alpha,\,1-1/\alpha\right)=\pi/\sin(\pi/\alpha)\). The corresponding momentum-space wavefunction is given by~\cite{2007JMP....48g2105D}
\begin{equation}
\phi(p)=\left[\frac{\hbar\sin(\pi/\alpha)}{\alpha-1}\left(-\frac{D_\alpha}{E}\right)^{1/\alpha}\right]^{1/2}
\frac{\alpha}{1-D_\alpha |p|^\alpha/E},
\end{equation}
where the normalization coefficient is fixed by our conventions. The Fourier transform of this function agrees exactly with the Fox \(H\)-function result of Ref.~\cite{2007JMP....48g2105D}, but is not in good agreement with the elementary expressions given in Ref.~\cite{2008JMP....49e2105D}. To ensure accuracy, we obtain the coordinate-space wavefunction via Eq.~\eqref{Fourier} rather than using the approximate expressions of Ref.~\cite{2008JMP....49e2105D}.

\begin{figure}[htbp]
    \centering
    \includegraphics[width=0.4\textwidth]{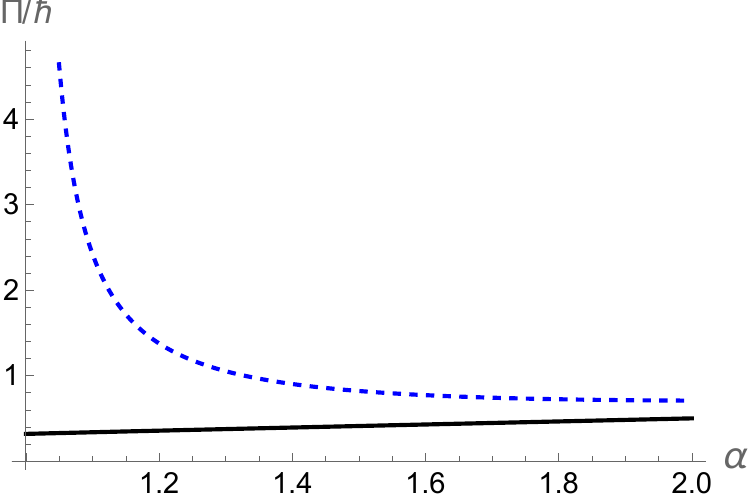}\\
    \caption{Position-momentum uncertainty product with \(\mu=\alpha^-\) for the bound state of the delta potential well (blue dashed curve) and for the L\'{e}vy wave packet~\eqref{Lup} (black solid line).}\label{fig:delwell}
\end{figure}

Since \(\phi(p)\) is even in \(p\), both \(\phi^*(p)\phi'(p)\) and \(p|\phi(p)|^2\) are odd functions. This immediately gives \(\langle x \rangle = \langle p \rangle = 0\) according to Eq.~\eqref{mean}. Again, the mean-\(\mu\) deviations of position and momentum reduce to absolute moments~\eqref{meanab}. For \(1<\mu<\alpha\le2\), the integral
\begin{equation}
\frac{1}{\pi\hbar}\int_0^\infty dp\,p^\mu|\phi(p)|^2
=
\frac{(\alpha-\mu-1)\sin\frac{\pi}{\alpha}}
{(\alpha-1)\sin\frac{\pi(\mu+1)}{\alpha}}
\left(-\frac{E}{D_\alpha}\right)^{\mu/\alpha}
\end{equation}
can be evaluated explicitly. Taking the limit \(\mu\to\alpha^-\), we obtain
\begin{equation}
\lim_{\mu\to\alpha^-}
\langle |\Delta p|^{\mu}\rangle^{1/\mu}
=
\left[
-\frac{E}{(\alpha-1)D_\alpha}
\right]^{1/\alpha}.
\end{equation}
The integral defining \(\langle |\Delta p|^{\mu}\rangle\) is evaluated numerically, and the resulting position-momentum uncertainty product is plotted in Fig.~\ref{fig:delwell}. As is evident, the revised bound is once again satisfied.

As a consistency check, one can take the \(\alpha\to2\) limit of our results and compare them with standard quantum mechanics. It is also interesting to note how the uncertainty product \(\Pi\) depends on the L\'{e}vy index \(\alpha\). As \(\alpha\) increases from \(1^+\) to \(2\), \(\Pi\) increases for the examined states in the infinite square potential well, but decreases for the bound state in the delta potential well.

\section{Discussion}\label{sect-disc}
This work revisits the decade-long debate on the fractional uncertainty relation and proposes a resolution. By examining the position-momentum uncertainty product of the L\'{e}vy wave packet analytically and numerically, we confirm that Laskin's original lower bound is violated near the boundary \(\alpha=1\), and find that this violation stems from an invalid estimate in the original derivation. Preserving the spirit of Laskin's proposal while correcting the calculation, we propose a revised bound that is nearly linear in the L\'{e}vy index \(\alpha\) across the full range \(1<\alpha\le2\). We further test the new bound against exact solutions of the fractional Schr\"{o}dinger equation for the infinite square potential well and the delta potential well. In both cases, the uncertainty products lie above the revised bound, confirming its validity.

A notable feature of the revised bound is its simplicity: it is almost linear in \(\alpha\), despite the intricate analytic structure of the underlying wave packet. This linearity suggests that the leading dependence on the L\'{e}vy index captures the essential physics, with higher-order corrections remaining small.

Several limitations and open questions remain. First, our tests are restricted to a limited set of exact solutions, and it would be desirable to verify the bound for other potentials, such as the finite square well and the Coulomb potential. Second, the revised bound is proposed on the basis of analytical and numerical studies of the L\'{e}vy wave packet; a rigorous proof from first principles would strengthen its status. Finally, experimental verification of fractional uncertainty relations remains challenging, but recent progress in simulating fractional dynamics in optical systems may offer new opportunities. We hope that the present work stimulates further theoretical and experimental investigations of uncertainty relations in fractional quantum mechanics.

\begin{acknowledgments}
This work is sponsored by the Natural Science Foundation of Shanghai (Grant No. 24ZR1419300) and the National Natural Science Foundation of China (Grant No. 12575060).
\end{acknowledgments}

\appendix
\section{Evaluation of coefficient functions~\eqref{Iin}}\label{app-coef}
The integrals in Eq.~\eqref{Iin} can be obtained from the single analytic function
\begin{equation}\label{G}
G(s,\zeta) = \int_{-\infty}^{\infty} du\; e^{iu\zeta}\, e^{-|u|}\, |u|^s,\qquad \operatorname{Re}(s) > -1,
\end{equation}
by evaluating the function and its first derivative with respect to \(s\) at specific values of \(s\). Concretely,
\begin{equation}
I_0(\zeta) = G(0,\zeta),\qquad I_1(\zeta) = -G'(1,\zeta).
\end{equation}

Splitting the integral \eqref{G} at \(u = 0\) and then substituting \(u \to -u\) in the second term yields
\begin{equation}
G(s,\zeta) = \int_0^{\infty} u^s e^{-u(1-i\zeta)} du + \int_0^{\infty} u^s e^{-u(1+i\zeta)} du.
\end{equation}
Using
\begin{equation}
\int_0^{\infty} u^s e^{-a u} du = \Gamma(s+1) a^{-(s+1)}
\end{equation}
for \(\operatorname{Re}(a)>0\), we obtain
\begin{equation}
G(s,\zeta) = \Gamma(s+1)\bigl[(1-i\zeta)^{-(s+1)} + (1+i\zeta)^{-(s+1)}\bigr].
\end{equation}
For \(s=0\), \(\Gamma(1)=1\), so
\begin{equation}\label{I0}
I_0(\zeta) = (1-i\zeta)^{-1} + (1+i\zeta)^{-1} = \frac{2}{1+\zeta^2}.
\end{equation}

To calculate \(I_1(\zeta)\), we differentiate \(G(s,\zeta)\) with respect to \(s\),
\begin{align}
\nonumber G'(s,\zeta) &= \Gamma'(s+1)\bigl[(1-i\zeta)^{-(s+1)}+(1+i\zeta)^{-(s+1)}\bigr] \\
\nonumber &\quad - \Gamma(s+1)\bigl[(1-i\zeta)^{-(s+1)}\ln(1-i\zeta) \\
&\quad + (1+i\zeta)^{-(s+1)}\ln(1+i\zeta)\bigr].
\end{align}
Setting \(s=1\) and using \(\Gamma(2)=1\), \(\Gamma'(2)=1-\gamma\), we obtain after some algebra
\begin{align}\label{I1}
\nonumber I_1(\zeta) &= -(1-\gamma)\bigl[(1-i\zeta)^{-2}+(1+i\zeta)^{-2}\bigr] \\
\nonumber &\quad + \bigl[(1-i\zeta)^{-2}\ln(1-i\zeta)+(1+i\zeta)^{-2}\ln(1+i\zeta)\bigr] \\
&= \frac{1-\zeta^2}{(1+\zeta^2)^2}\bigl[\ln(1+\zeta^2) - 2(1-\gamma)\bigr] + \frac{4\zeta\arctan\zeta}{(1+\zeta^2)^2}.
\end{align}
Both \(I_0\) and \(I_1\) are real and even in \(\zeta\) because the Fourier transforms of even functions are real and even.


\begin{thebibliography}{99}
\bibitem[Konishi, Paffuti, and Provero(1990)]{1990PhLB..234..276K}Konishi, K., Paffuti, G., and Provero, P.: 1990, {\it Physics Letters B} {\bf 234}, 276. 
\bibitem[Ellis, Mavromatos, and Nanopoulos(1992)]{1992PhLB..293...37E}Ellis, J., Mavromatos, N.E., and Nanopoulos, D.V.: 1992, {\it Physics Letters B} {\bf 293}, 37. 
\bibitem[Bender and Boettcher(1998)]{1998PhRvL..80.5243B}Bender, C.M. and Boettcher, S.: 1998, {\it Physical Review Letters} {\bf 80}, 5243. 
\bibitem[Laskin(2000)]{2000PhLA..268..298L}Laskin, N.: 2000, {\it Physics Letters A} {\bf 268}, 298. 
\bibitem[Kaplan and Rajendran(2022)]{2022PhRvD.105e5002K}Kaplan, D.E. and Rajendran, S.: 2022, {\it Physical Review D} {\bf 105}, 055002. 
\bibitem[Laskin(2000)]{2000Chaos..10..780L}Laskin, N.: 2000, {\it Chaos} {\bf 10}, 780. 
\bibitem[Laskin(2000)]{2000PhRvE..62.3135L}Laskin, N.: 2000, {\it Physical Review E} {\bf 62}, 3135. 
\bibitem[Laskin(2002)]{2002PhRvE..66e6108L}Laskin, N.: 2002, {\it Physical Review E} {\bf 66}, 056108. 
\bibitem[Guo and Xu(2006)]{2006JMP....47h2104G}Guo, X. and Xu, M.: 2006, {\it Journal of Mathematical Physics} {\bf 47}, 082104. 
\bibitem[Dong and Xu(2007)]{2007JMP....48g2105D}Dong, J. and Xu, M.: 2007, {\it Journal of Mathematical Physics} {\bf 48}, 072105. 
\bibitem[Dong and Xu(2008)]{2008JMP....49e2105D}Dong, J. and Xu, M.: 2008, {\it Journal of Mathematical Physics} {\bf 49}, 052105. 
\bibitem[de Oliveira, Costa, and Vaz(2010)]{2010JMP....51l3517D}de Oliveira, E.C., Costa, F.S., and Vaz, J.: 2010, {\it Journal of Mathematical Physics} {\bf 51}, 123517. 
\bibitem[Capelas de Oliveira and Vaz(2011)]{2011JPhA...44r5303C}Capelas de Oliveira, E. and Vaz, J.: 2011, {\it Journal of Physics A Mathematical General} {\bf 44}, 185303. 
\bibitem[Dong and Lu(2021)]{2021JMP....62c2104D}Dong, J. and Lu, Y.: 2021, {\it Journal of Mathematical Physics} {\bf 62}, 032104. 
\bibitem[Wei(2016)]{2016PhRvE..93f6103W}Wei, Y.: 2016, {\it Physical Review E} {\bf 93}, 066103. 
\bibitem[Laskin(2016)]{2016PhRvE..93f6104L}Laskin, N.: 2016, {\it Physical Review E} {\bf 93}, 066104. 
\end{thebibliography}
\end{document}